\documentstyle[11pt,aaspp4]{article}

\def\lax    {\ifmmode{_<\atop^{\sim}}\else{${_<\atop^{\sim}}$}\fi}
\def\gax    {\ifmmode{_>\atop^{\sim}}\else{${_>\atop^{\sim}}$}\fi}

\newcommand\op{{\rm cm^{2} \ g^{-1}}}

\received{ }
\accepted{ }

\slugcomment{To appear in The Astrophysical Journal (1999 November 10
issue)
}
\lefthead{Osorio et al.}
\righthead{Model of  Hot Molecular Cores}
\begin{document}

\title{ HOT MOLECULAR CORES AND THE FORMATION OF MASSIVE STARS}

\author{Mayra Osorio, Susana Lizano}

\affil{Instituto de Astronom\'\i a, Unidad Morelia, UNAM, J.J Tablada
1006,\\ Col. Lomas de Santa Mar\'\i a,  58090 Morelia, Michoac\'an, MEXICO}
\and
\author{Paola D'Alessio}

\affil{Instituto de Astronom\'\i a, UNAM, Apdo Postal 70-264, Cd. 
Universitaria,\\ 04510 M\'exico D. F.,
MEXICO}

\begin{abstract}

It has been proposed that some hot molecular cores (HMCs) harbor a young
embedded massive star, which heats an infalling envelope and accretes
mass at a rate high enough to ``choke off'' an incipient HII region. This
class of HMCs would mark the youngest phase known of massive star
formation. In order to test this hypothesis, we model this type of object
calculating the radiative transfer through a spherically symmetric dusty
envelope infalling onto a central OB star, with accretion rates from $\dot
M =$ 6~$\times$ 10$^{-4}$ to 10$^{-3}$ M$_{\odot}$ yr$^{-1}$. The dust 
thermal spectrum from infrared to radio wavelengths is derived and is 
compared with the observed fluxes of several hot cores which may be 
internally heated. We find that the data are best fitted using an envelope 
with the density distribution resulting from the collapse of a singular 
logatropic sphere, instead of that of a singular isothermal sphere. We 
conclude that several of these sources may be undergoing an intense accretion 
phase and find in all the cases that the accretion luminosity exceeds the 
stellar luminosity. We discuss the implications of this phase on the 
formation of massive stars.

\end{abstract}


\section{INTRODUCTION}

One of the most interesting results of the past few years in the study of
massive star formation is the discovery of Hot Molecular Cores (HMCs) which 
are small ($\la 0.1$~pc), dense ($n\sim10^6$ - $10^8$~cm$^{-3}$), hot 
(T $\sim$ 100 - 300 K) and dark ($A_v \sim 10^3$ mag)  molecular clumps. 
They are generally found in the proximity of ultra-compact HII (UCHII) 
regions, and are probably directly related to the formation of high mass 
stars. The best known example is the Orion-KL hot core (e.g., Genzel \& 
Stutzki 1989;  Kaufman, Hollenbach \& Tielens 1998 and references therein) 
but several other cores have been observed with qualitatively similar 
characteristics: the core near G31.41+0.31 (Cesaroni et al. 1994b), 
G9.62+0.19F (Hofner et al. 1996), G34.24+0.13MM (Hunter et al. 1998, 
hereafter H98), W3(H$_2$O) (Turner \& Welch 1984), and IRAS 23385+6053  
(Molinari et al. 1998; hereafter M98) among others. These sources have 
{\em not} been detected as free-free emitters at centimeter wavelengths; 
rather, their presence is established, and their physical properties 
determined through milli\-meter-wave and molecular line observations 
(see the review of Kurtz et al. 1999).

It is not yet understood what heats the gas in HMCs to a temperature
higher than $100$ K. In some cases, the heating source might be a nearby
star, facing the clump. This seems to be the case of the warm  ammonia clumps 
associated with the source  G61.48+0.09 (G\'omez et al. 1995). Alternatively, 
Walmsley (1995) proposed that  the heating source  could be a recently formed 
OB-type star (or stars) inside the core undergoing an intense accretion 
phase. Following Yorke (1984), he suggested that high mass accretion rates of 
the infalling material could quench the development of an UCHII region, so 
that the free-free emission from the ionized material would be undetectable 
at centimeter wavelengths. Observationally, Cesaroni et al. (1994a) found 
that several HMCs seen in NH$_3$(4,4) are coincident with groups of water 
masers, which are known signposts of massive star formation. Thus, these 
sources are of particular interest because they may represent the youngest 
phase yet observed in the life of a massive star, and they could help us 
understand the process of high-mass star formation. 

Kaufman et al. (1998) modeled the temperature distributions of internally and 
externally heated molecular cores and calculated the column density of 
hot dust and gas inside the cores. They considered constant density 
distributions and collapsing density distributions ($\rho \propto r^{-3/2}$) 
for the cores. In the latter case, they did not discuss explicitly the effect 
of the accretion luminosity in the core heating. They applied their models to 
the Orion hot core with mass $M_{\rm core} \sim 15~ M_\odot$, and size 
$r_{\rm core} \sim 0.01$ pc (i.e. a column density of  $N(H_2) > 5 \times 
10^{24}$ cm$^{-2}$). These authors concluded that an internal energy source 
is necessary to heat up the entire column density of the core to the observed 
temperatures, $ T > 200 $ K, since the nearest known  source (Radio 
Source ``I")  is not luminous enough to externally illuminate and produce the 
observed high temperatures of the gas and dust. 

In this paper, we calculate the intrinsic dust thermal spectrum from 
infrared to radio wavelengths for a simple theoretical construct of the 
temperature profile of a HMC: a central massive star undergoing spherical 
accretion of a free falling envelope of gas and dust. The central heating 
source has two components, the stellar luminosity and the accretion 
luminosity. 

We compare the spectrum of our model with the observed fluxes in several 
sources to determine the physical size of the HMC, the mass accretion rate of 
the envelope, and the spectral type of the central star. We present in $\S 2$ 
the parameters and the assumptions made in our model, in $\S 3$ the model 
results, in $\S4$ a comparison with the available observations, in $\S 5$ a 
discussion our results, and in $\S 6$ the conclusions.

\section{THE MODEL}

Inspired by the scenario proposed by Walmsley (1995), we model a HMC as
an envelope of gas and dust that is freely falling onto a recently formed 
massive central star. The young star is embedded within the dense core 
and interacts with it through its radiation, heating the core from inside. 

We assume that the system consists of a central heating source, with a 
total bolometric luminosity $L_{\rm core}$, surrounded by a spherically 
symmetric envelope, with inner radius $R_{\rm d}$ and outer radius 
$R_{\rm core}$.  The inner radius of the envelope is set by the dust 
destruction radius (see $\S 2.4$ ).  The luminosity of the heating source, 
$L_{\rm core}=L_*+L_{\rm acc}$, has two components, the luminosity of the
central star (considered to be a ZAMS massive star), $L_*$, and the accretion
luminosity, $L_{\rm acc} = {G M_* \dot {M}/R_*}$, where $G$ is the 
gravitational constant, $M_*$ is the stellar mass, $\dot {M}$  is the mass 
accretion rate, and $R_*$ is the stellar radius.

The effect of an energetic  stellar wind is not taken into account. In 
spherical symmetry this is justified as long as the mass accretion rate is 
larger that the wind mass loss rate, $\dot M_{\rm acc} > \dot M_w$, since the 
wind terminal speed is of the order of the free-fall velocity. In this case 
the ram pressure of the accretion flow will prevent the wind from escaping 
from the stellar surface. More likely, the angular momentum of the infalling 
envelope will deviate the flow from spherical accretion and the material will 
be deposited in a circumstellar disk around the star. In this case, the 
stellar winds could escape more easily through the poles. Our models will 
still be valid if the stellar winds are ejected in narrow bipolar cones, 
coexisting with the accreting envelope as in the case of low mass proto-stars 
(Adams, Lada $\&$ Shu 1987), as long as these cones of missing material 
represent only a small fraction of the total core material and provided the 
axis of the bipolar outflow is not close to the line-of-sight. Also, the 
deviations of the envelope density profile from  spherical symmetry  will 
be important only within the centrifugal radius (Adams \& Shu 1986). This
region does not make a significant contribution to the emergent flux 
as long as it is small compared to the size of the core. Since this is 
the case expected for our models, the predicted spectra will be appropriate.

To calculate the emergent dust thermal spectrum, we ignore the interaction 
of radiation from the central object with envelope matter located inside 
$R_{\rm d}$. We calculate the flux received from the source by an observer 
at a distance $D$ by  integrating the equation of radiative transfer  along 
rays for different impact parameters (e.g. Mihalas 1978).  We take the volume 
emissivity of the dust to equal the  LTE value 
$\rho \kappa_{\nu} B_{\nu}(T)$, where $\rho$ is the matter density, 
$\kappa_{\nu}$ is the monochromatic absorption opacity per unit mass, 
$B_{\nu}$ is the Planck function, and $T$ is the local dust temperature. In 
order to solve the transfer equation, a knowledge of $\rho$, ${\kappa_{\nu}}$ 
and $T$ is needed. 

\subsection{Density Distribution}

Two  different types of density distributions are considered: one resulting 
from the collapse of the singular isothermal sphere (SIS; Shu 1977), and the 
other from the collapse of the singular logatropic sphere (SLS; McLaughlin 
\& Pudritz 1997). The logatropic equation of state, 
$P = P_0 \ln (\rho/\rho_0)$, has been invoked to explain the linewidth-size 
relation  observed in molecular clouds (Lizano \& Shu 1989; Myers \& Fuller 
1992; McLaughlin \& Pudritz 1996) since the sound speed, $c_s$, given by 
$c_s^2 = {d P / d \rho} = {P_0 / \rho}$, behaves like the observed velocity 
dispersion in these clouds. In the above equations, $P_0$ has a 
constant value and $\rho_0$ is an arbitrary reference density. 

The logatropic and isothermal  collapses occur in the same general fashion:
an expansion wave moves outward into a cloud at rest and the gas behind the 
wave falls into the central proto-star. The infalling matter close to the 
center has free-fall density and velocity profiles given by 
$\rho = \dot M (32 \pi^2 G M_*)^{-1/2} ~r^{-3/2}$, and 
$v= - (2 G M_*)^{1/2} r^{-1/2}$, respectively, while in the outer region 
the density profile tends towards the hydrostatic equilibrium configuration: 
$\rho(r) =  a^2 (2 \pi G)^{-1} r^{-2}$ for the SIS, where $a$ is the 
isothermal sound speed, and $\rho(r) = (P_0/ 2\pi G)^{1/2}r^{-1}$ for the SLS.
Nevertheless, there are important differences in the behavior of both types 
of collapsing clouds. The mass accretion rate is constant for the SIS 
collapse: $\dot M^i =m^i_0 a^3 / G$, where the  reduced mass is
$m^i_0$ = 0.975. For the SLS collapse, however, the mass 
accretion rate is a steep function of time given by
\begin{equation}
\dot M = {(2 \pi G P_0)^{3/2} \over 4 G} m_0 t^3,
\end{equation}
and  the reduced mass is $m_0 = 0.0302$. 

For a given stellar mass, $M_*$, and present mass accretion rate, $\dot M$, 
the age of the system in the SIS is given by $t^i_{\rm age} = M_*/ \dot M$, 
while it is longer for the SLS
\begin{equation}
t_{\rm age} = 4 {M_* \over \dot M}  = 4 t^i_{\rm age}.
\end{equation}

Furthermore, for a given central star, the mass of the infalling envelope is 
given by
\begin{equation}
M_{\rm env} = M_* \left( {m(x) \over m_0} -1 \right),
\end{equation}
where $m(x)$ is the dimensionless mass function of the dimensionless variable 
$x$. For the SIS, $x=r/at$, where $r$ is the distance to the central star, 
while $x = {4 (2 \pi G P_0)^{-1/2} } { r t^{-2}}$, for the SLS. For 
$t_{\rm age}$, the location of the expansion wave, $r_{\rm ew}$, occurs 
at $x=1$. Then, one has the infalling density distribution within $r_{\rm ew}$ 
and the hydrostatic density distribution outside this radius. In particular, 
at the expansion wave, $m(1)=2$ for the SIS and $m(1)=1$ for the SLS. Thus, 
at a given time, within the expansion wave, the SLS has only 3 $\% $ of the 
mass in the star and the rest is contained in the envelope. In contrast, 
for the SIS,  $m^i_0=0.975$, so about half of the mass is in the star and 
half in the envelope. Also, eqns. (2) and (3) imply that, for a given $M_*$ 
and $\dot M$, the  isothermal envelope has less mass than the logatropic 
envelope. For simplicity, we assume that the density distribution is not 
affected by effects of the finite core size. 

\subsection{Dust Opacity}

We assume that the dust in the core is a mixture of 4 types of grains: 
graphite, silicates, iron and water ice, with optical constants and 
abundances taken from D'Alessio (1996; see references therein), and a 
MRN-type size distribution (Mathis, Rumpl $\&$ Nordsieck 1977). 
Water ice only contributes to the opacity for T $<$ 100 K. For higher 
temperature we assume it has sublimated (Sandford \& Allamandola 1993).
 Since little is known about grain opacities in the sub mm range, we adopt a 
power law of the form, $\kappa_{\lambda} \propto\lambda ^{-\beta}$, with 
$1 \leq \beta \leq 2$ for $\lambda$ $\ge$ 200 $\mu$m. With these 
monochromatic opacities, the Rosseland mean opacity, $\chi_{\rm R}$, 
and the Planck mean opacity, $\kappa_{\rm P}$, are calculated in the 
wavelength range $0.1< (\lambda/\mu m) < 10^5$. In this range, the 
monochromatic opacities for the low and high temperatures in the envelope 
($\sim$ 20 - 2000 K), are well represented.

\subsection{Temperature Profile of the Dust Envelope}

Kenyon, Calvet \& Hartmann (1993) made detailed models of the temperature 
structure and emission of dusty infalling envelopes around low mass 
proto-stars. They proposed a simplified calculation of the temperature 
profile which we have adopted to calculate the temperature distribution in 
our model. The basic idea of this approximate treatment considers radiative 
equilibrium for the outer optically thin dusty envelope and assumes the 
standard diffusion approximation in the inner optically thick region. The 
radius that divides the optically thin and thick  regions is called the 
``photospheric radius", $R_{\rm ph}$. The total luminosity is 
conserved  at this radius, thus, $L_{\rm core} = 4 \pi R_{\rm ph}^2 
\sigma T_{\rm ph}^4$, where $\sigma$ is Stefan-Boltzmann constant and 
$T_{\rm ph}=T(R_{\rm ph})$. We also require that $\tau_{\rm R}(T_{\rm ph}) 
= 2/3$, where $\tau_{\rm R} (T_{\rm ph})$ is the  Rosseland mean 
optical depth weighted by the Planck function evaluated at $T_{\rm ph}$. 
Both, the optical depth and the luminosity conservation conditions are used 
to determine $R_{\rm ph}$ and $T_{\rm ph}$. Unlike the SIS collapse density  
distribution, in the collapse of a  SLS the photospheric radius  changes for 
different external radii of the envelope because the external part of the SLS 
envelope has an appreciable optical depth.

The equation for radiative equilibrium is
\begin{equation}
\int_{0}^{\infty} \kappa_{\nu}B_{\nu}[T(r)]d{\nu} = \int_{0}^{\infty}
\kappa_{\nu}J_{\nu}d{\nu}.
\end{equation}
For the optically thin outer region,  the  mean intensity,  $J_{\nu}(r)$, 
is approximated as $J_{\nu}(r)=W(r)B_{\nu}(T_{\rm ph})$, where $W(r)$ is the 
dilution factor: 
$ W(r) = {1 \over 2} ( 1 - [1 - (R_{\rm ph}/r)^2]^{1 \over 2} )$.
Thus, one accounts for the geometric dilution of the radiation field of 
the envelope's photosphere but not the exponential attenuation of this 
emission which, by assumption, is less than e$^{-{2/3}}$. Therefore, we 
consider that the grains in this region are heated only by the energy from 
the photosphere and we neglect the heating by the diffuse field from the 
optically thin region. Using the Planck mean opacities, the thermal balance 
equation for the dust becomes  
\begin{equation}
T^4 \kappa_{\rm P} (T) = W(r) T_{\rm ph}^4 \kappa_{\rm P} (T_{\rm ph}).
\end{equation}
where the Planck mean opacity, $\kappa_{\rm P}(T)$, is evaluated at the local 
dust temperature, $T$, and $\kappa_{\rm P} (T_{\rm ph})$ is evaluated at 
$T_{\rm ph}$. Once the photospheric temperature is known, the implicit 
eqn. (5) determines the temperature distribution, $T(r)$ in the outer region.

The innermost regions of the accreting dust envelope (inside the photospheric 
radius) are optically thick. In such regions the temperature gradient is 
determined by the standard diffusion approximation 
\begin{equation}  
L_{\rm core} = - { {64 \pi \sigma r^2 T^3 \over 3{ \chi_{\rm R} \rho}}
{dT \over dr}}.
\end{equation}

For a given density  $\rho(r)$ and Rosseland mean opacity $\chi_{\rm R}(T)$, 
one solves this equation for the temperature distribution $T(r)$, using a 
Runge-Kutta integration technique subject to the boundary condition 
$T(R_{\rm ph})=T_{\rm ph}$. The approximation used here to obtain the 
temperature distribution is a very simplified approach since a full angular 
and frequency-dependent radiative transfer treatment is beyond the scope of 
this paper. 

\subsection{Dust Destruction Radius}

The dust is sublimated for temperatures higher than $\sim$ 1200 K (e.g. Adams 
\& Shu 1985), producing an inner dust free cavity. The dust destruction 
radius is calculated following the evolution of an average sized dust 
particle ($a_{g}$ = 0.1 $\mu$m) until it disappears, neglecting particle 
accretion onto the grains. The equation for the grain radius variation is 
given by 
\begin{equation}
{{\Big({d{a_{g}} \over dt}\Big)}_{\rm sub}} ={{-1\over {{ \rho_g}}}}
{{\sqrt{3 {\mu _g} {m_H} \over{16k T(r)}}} }{ P_{\rm vap}}, 
\end{equation}
where  $\rho_{\rm g}$ is the grain mass density, $\mu_{\rm g}$ is the 
mean molecular weight of the gas and dust material, $m_{\rm H}$ is the 
hydrogen mass and  ${ P_{\rm vap}}$ is the vapor pressure, taken
from Lamy (1974). $P_{\rm vap}$ depends on the dust grain composition, 
which we have assumed to be pure silicate. With this approximation one 
obtains $T(R_{\rm d})$ $\sim$ 1200 K, where the dust destruction 
radius, $R_{\rm d}$, is of order tens of AUs (see Tables 1a, 1b and 3). In a 
more rigorous analysis one should calculate the destruction radius of each 
one of the components and sizes of the dust grain mixture.

\subsection{Central Stars}

To determine the total luminosity, we must specify not only the stellar
mass and mass accretion rate but also the radius and luminosity of the
central star.  It is unclear whether a star formed under very high
accretion rates, will have a normal ZAMS radius. In fact, Adams \& Shu
(1985), using the results of Stahler, Shu \& Taam (1980) for the formation
of low mass stars, found that the stellar radius is an increasing function
of accretion rate:  $R_* \propto \dot M^{0.33}$. Furthermore, Stahler,
Palla \& Salpeter (1986)  studying the formation of primordial massive
stars, found the same type of behavior: $R_* \propto \dot M^{0.41}$. In
addition, Beech $\&$ Mitalas (1994)  and Bernasconi $\&$ Maeder (1996)
have studied the formation of massive stars with solar abundances and
under constant accretion rates. These authors do not discuss enlarged
radii, maybe because they studied only accretion rates $< 10^{-4}
M_{\odot}$ yr$^{-1}$. It is then quite probable that OB stars, formed
under intense accretion flows inside HMCs, will have radii larger than the
ZAMS values since they have not had time to get rid of excess internal
energy. We consider two possibilities: stars with ZAMS radii of Thompson
(1984) (see also Vacca, Garmany, $\&$ Shull 1996), and stars with stellar
radii $R_* = 10^{12}$ cm. This last value for the radius is a factor of
$\sim$ 2-5 larger than the ZAMS radii for the central B0-B3 stars
considered in our models and is within the range of radii found by the
authors mentioned above. Finally, we adopt the ZAMS luminosities of
Thompson (1984; see also Vacca, Garmany, \& shull 1996), even though, 
the luminosities of massive stars forming under intense accretion rates are 
also uncertain. 

\section{MODEL RESULTS}

This section  presents the results obtained for a set of models with 
different values of the mass accretion rate, the external radius of the 
envelope, and the spectral type of a ZAMS central star.
We discuss the effects in the spectrum, when the above parameters are 
changed, assuming the collapse density distribution of the SLS and
SIS. For both cases, a distance of 4.9 kpc 
was assumed for the model source, namely the distance to IRAS 23385+6053, 
which is one of the studied HMCs.

The summary of the physical properties of the SLS models is given in 
Table 1a and its  general trends are shown in Figure 1a. The top panel of 
Figure 1a shows the spectra 
obtained for three different values of the mass accretion rate (${\dot
M}=1.5\times10^{-3}$, $4.4\times10^{-4}$ and $8.1\times10^{-5}~
M_{\odot}$~yr$^{-1}$), for a B0 central star with a radius $R_* = 4
\times 10^{11}$ cm (Thompson 1984), and an envelope with an external 
radius $R_{\rm core}=0.1$ pc. As seen in the figure, the peak flux density  
increases and shifts to longer wavelengths as the accretion rate increases. 
This behavior 
can be explained as a result of the increase of the optical depth in the 
envelope, because of the overall density increase for higher accretion rates. 
As a consequence of the increase in opacity, the flux tends to be 
redistributed towards longer wavelengths, where the envelope is optically 
thinner, producing a shift of the peak of the spectrum. This also has the 
effect of increasing the depth of the silicate absorption feature at 10 
$\mu$m. The total luminosity, $L_{\rm core}$, increases with the mass 
accretion rate because of the increase in the accretion luminosity. This 
results in higher values of the peak flux density. Table 1a shows that the 
photospheric radius increases and its temperature decreases with increasing 
mass accretion rates.Again this is easily understood as a result of the 
overall density  increase in the envelope that moves the photospheric radius 
to outer (and  thus cooler) regions, as the accretion rate increases. Table 1a 
also indicates that the dust destruction radius, $R_{\rm d}$, increases with  
the accretion rate, since the central luminosity, $L_{\rm core}$, increases.

The middle panel of Figure 1a shows the spectra for different spectral types 
of the central star (O7, B0 and B3), for an envelope with external radius of 
0.1 pc, and an accretion rate of $4.4\times 10^{-4}~M_{\odot}$ yr$^{-1}$. 
The peak flux density of the spectrum shifts to longer wavelengths for
later spectral types. This happens because, for the models presented in this
figure, $R_{\rm core} < r_{\rm ew}$ (see column 7 in Table 1a), thus, the 
mass of the envelope is given 
by eqn. (3), where $m(x) \propto x^{1.84}$, and 
$x \propto M_*^{-1} \dot M^{2/3}$. Hence, the envelope mass is inversely 
proportional to the stellar 
mass, i.e. $M_{\rm env} \propto M_*^{-0.84}$ (see last column in Table 1a). 
Therefore, the opacity of the 
envelope increases for later spectral types and the emission is redistributed 
towards longer wavelengths.

Finally, the bottom panel of Figure 1a illustrates the trends of the spectra 
for different external radii of the envelope ($R_{\rm core}$ = 0.15, 0.10 and 
0.05 pc), for a central star with spectral type  B0 and an 
accretion rate of $4.4\times 10^{-4}~M_{\odot}$ yr$^{-1}$. As seen in the 
figure, the flux at long wavelengths increases slightly, while it 
decreases slightly at short wavelengths, as the external radius increases. 
This effect can be also understood as a consequence of the redistribution of 
the flux density  to longer wavelengths, because of the increase of the 
optical depth for larger external radii. Note that, the photospheric radius 
increase with increasing  external radii, as 
discussed in $\S$ 2.3.

The summary of the physical properties of the SIS models is given in 
Table 1b and its general trends are shown in Fig 1b. The SIS models in each 
panel were computed with the same parameters that the SLS models (see columns 
1 and 2 in Tables 1a and 1b). As one can see from Figures 1a and 1b, that the 
general behavior of the SIS models in each panel is the same the SLS models.
However, one can also see that the spectra resulting of the SIS models have 
smaller millimeter fluxes and greater infrared fluxes than the SLS. This 
happens because the mass of the SIS envelope, 
M$_{\rm env}={{\dot M}{(18GM_*)^{-1/2}}}R_{\rm core}^{3/2}$, is always 
smaller than that of the SLS envelope (see last columns in Tables 1a and 1b). 
Thus, the SIS envelope is optically thinner and  emits most of the 
radiation at shorter wavelengths than the SLS envelopes.

In summary, the spectrum at all frequencies is very sensitive to the mass
accretion rate; infrared flux densities are sensitive to the spectral type
of the central star, and the value of the external radius has only a small
effect on the spectrum for the parameters explored in Figures 1a and 1b.

\section{COMPARISON OF THE MODEL WITH OBSERVATIONS}

To date, about 20 candidate HMCs have been proposed (Kurtz et al. 1999).
In general, these objects are associated with nearby or embedded FIR sources 
and/or UCHII regions. Thus, it is very likely that the observed fluxes are 
contaminated by these nearby sources and overestimate the actual contribution 
of the HMC, making it difficult to fit  the intrinsic spectrum. For this 
reason, we selected relatively isolated objects, and among them, those with 
the most complete data sets available in order to constrain, as much as 
possible, the parameters of our  model. With these restrictions, we chose the 
sources: G34.24+0.13MM, W3(H$_2$O), Orion hot core and IRAS 23385+6053 as a 
representative sample to test the  model.

In order to match the sources, fits using both the SLS and SIS density 
distributions have been considered. Figure 2 shows the observed fluxes of 
the source G34.24+0.13MM (H98) and the results of three model spectra for 
HMCs with $R_{\rm core}$ = 0.07 pc and a central B3 star with 10 $M_{\odot}$. 
The dotted line corresponds to the SIS collapse density distribution with 
${\dot M} = 1.0 \times 10^{-3}~M_{\odot}~{\rm yr}^{-1}$, while the dot-dashed 
line corresponds to the SLS collapse density distribution with 
${\dot M} = 6.5 \times 10^{-4}~M_{\odot}~ {\rm yr}^{-1}$. For both models we 
used a ZAMS stellar radius, $R_*=2.0 \times 10^{11}$ cm (Thompson 1984). In 
order to match the observed fluxes at mm wavelengths, massive envelopes with 
high accretion rates are required. In particular, for a given mass accretion 
rate, the envelope is less massive for the SIS than the SLS, so one requires 
a higher mass accretion rate, producing a flux density at $20~\mu$m several 
orders of magnitude higher than the upper limit on the flux at this wavelength 
(see discussion in $\S$ 3). The logatropic collapse model, on the other hand, 
has a smaller excess emission at 20 $\mu$m which can be attenuated invoking a 
high external extinction to the HMC, $A_V \sim 100$ mag. This value of the 
external extinction is somewhat high, even though HMCs are found inside dense 
regions in molecular clouds. More generally accepted values are 
$A_V \lesssim 30$ mag (e.g. Testi et al 1998). 

The solid line in Figure 2 shows the spectrum of the SLS model discussed 
above, but for a star of radius  $R_* = 10^{12}$ cm, as discussed in $\S$ 
2.5. Clearly, the excess of emission at 20 $\mu$m is reduced, and now fits 
the constraint at $20~ \mu$m, with  no external extinction required, since 
the accretion luminosity decreases for this large stellar radius. Note that 
if this large stellar radius were used for the SIS density distribution, one 
still requires $A_V > 100$ mag to attenuate the excess emission at $20~\mu$m.

A similar result is found for the remaining sources, so, for these, we 
present only models for a SLS collapse distribution. In addition, the central 
stars are assumed to have radii $R_* = 10^{12}$ cm, and any external 
extinction is neglected. It is likely that in reality, a combination of some 
external extinction ($A_V \lesssim 30$ mag) plus a star with a radius larger 
than its ZAMS value, could be responsible for the observed spectrum. Finally, 
similar trends, as those shown in Figure 1, are found for these type of 
central stars with large radii.

\subsection{G34.24 + 0.13MM} 

The source G34.24 + 0.13MM discovered by Hunter et al. (1998), is located 
$84''$ (1.5 pc) southeast of the UCHII region G34.26 + 0.15, within the 
G34.3 + 0.2 HII region complex, which is at a distance of 3.7 kpc (Kuchar 
$\&$ Bania 1994). H98 identified G34.24 + 0.13 MM as a massive proto-stellar 
object, which coincides with a methanol maser. This source has  upper 
limits on the centimeter continuum emission of S$_{\nu} \lesssim$ 0.8 mJy at 
15 GHz, and S$_{\nu} \lesssim$ 0.6 mJy at 8.3 GHz (Goss 1999, personal 
communication). From observations at 1.3 mm and 2.7 mm, H98 measured an 
angular diameter of $2''$ (0.037 pc). They also obtained near-infrared images 
and photometry at 10 and 20 $\mu$m,  but no infrared source was detected 
coincident with the millimeter position. H98 estimated a total luminosity in 
the range of 1600 - 6300 L$_{\odot}$. These authors argue that because of the 
high luminosity and the lack of cm compact radio continuum emission the core 
probably contains a deeply embedded proto-B star, since there is no nearby 
source that could provide the observed heating.

We fitted the observed spectrum of this source with a  model requiring a 
central B3 star, ${\dot M}= 6.5 \times 10^{-4}~M_{\odot}~ {\rm yr}^{-1}$, 
$R_{\rm core} = 0.07$  pc, and an opacity index $\beta$ = 2 (see Table 2). 
Earlier spectral types than B3 are too luminous and are incompatible
with the observed mid-infrared emission (see trends of the models in the 
middle-panel of Figure 1a). Accretion rates smaller than 
the value ${\dot M}= 6.5 \times 10^{-4}~M_{\odot}~ {\rm yr}^{-1}$ do not 
reproduce the observed millimeter emission (see the trends of the models in 
the top-panel of Figure 1a). The top-left panel of Figure 3 shows the 
spectrum. Table 3 lists the physical properties of the source derived with 
this model: the dust destruction front, $R_{\rm d}$; the photospheric radius, 
$R_{\rm ph}$; the photospheric temperature, $T_{\rm ph}$; the molecular 
volume density at the external radius, $n(R_{\rm core})$; the total 
luminosity, $L_{\rm core}$; the fraction of the accretion luminosity, to the 
total luminosity, ${L_{\rm acc}}/L_{\rm core}$; the core mass within 
$R_{\rm core}$, $M_{\rm core}$; the total column density $N(H_2)$; the 
present age of the core given by eqn. (2); and the expansion wave radius. 

The required central B3 star is late enough so that it will not develop an 
HII region detectable from 3.7 kpc. Even if the star were earlier than B3, 
one would not expect centimeter radio emission since the critical mass 
accretion rate that would quench the development of an UCHII region (see 
Walmsley 1995, eqn. [1]) is only 
${\dot M}_{\rm crit}=3.20 \times 10^{-8}~M_{\odot}~{\rm yr}^{-1}$, 
considerably smaller than that implied by our model (see Table 2).

\subsection{ W3(H$_2$O)}

About 6$''$ (0.06 pc) east of the UCHII region W3(OH) located at a
distance of 2.2 kpc (Humphreys 1978), a luminous embedded young star was
discovered near a complex of H$_2$O masers by Turner $\&$ Welch (1984; 
hereafter the TW object or W3(H$_2$O)). This object is manifest by a warm,
dense concentration of molecular material (Turner \& Welch 1984; Wilson,
Gaume \& Johnston 1993; Wink et al. 1994), a bipolar outflow traced by
VLBI proper motions of H$_2$O maser spots (Alcolea et al. 1992) emanating
from a small, elongated synchrotron emission source (Reid et al. 1995; 
Wilner, Reid $\&$ Menten 1999). Additionally, Wilner, Welch $\&$ Forster
(1995), and Wyrowski et al. (1997) detected a compact ($\sim2''=0.02$ pc)
mm continuum source, that was interpreted as thermal dust continuum
emission from the W3(H$_2$O) hot core. Recent molecular line and continuum
interferometric observations at 1.4 mm, reaching subarcsecond angular
resolution (Wyrowski et al.  1999), indicate that the continuum emission
from W3(H$_2$O) splits into three components. The eastern component
(component A in the nomenclature of Wyrowski et al.  1999) coincides with
the source of the maser outflow and synchrotron emission, and is
associated with a 200 K temperature peak, derived from molecular line
transitions. Nevertheless, since our models are spherically symmetric, 
we model the spectrum of the compact source as observed by Wilner et al.
(1995) and  Wyrowski et al. (1997) with lower spatial resolution.

We fit the spectrum of this source with a model requiring a central B0
star, ${\dot M}= 1.2 \times 10^{-3}~ M_{\odot}~ {\rm yr}^{-1}$, $R_{\rm core} =
0.05$ pc, and an opacity index $\beta=1.6$ (see Table 2). Table 3 lists the
physical properties of this model and the resulting spectrum is shown in the
upper right panel of Figure 3. Because of the few data points available for 
this source, the  model fit was not unique, and we selected the model 
solution with the spectral type of the central star and the index of the 
opacity law in accordance with those inferred by Wyrowski et al. (1997, 
1999). High angular resolution observations at mid- and 
far-infrared wavelengths are necessary to constrain the total luminosity of 
this source, and  the spectral type of the central star.

\subsection{Orion hot core }

Because of its proximity ($D=480$ pc; Genzel et al. 1981), the most
 studied region of high-mass star formation is the Orion
molecular cloud (OMC-1). Line studies of the central portion of OMC-1
(i.e., the Kleinmann-Low nebula) showed that it was separated into a
number of velocity components that were subsequently associated with
several spatial components as very high-resolution mapping became
available (see Wright et al. 1996 and references therein). Prominent at
the center of this region is a dense condensation of gas and dust known as
the ``Orion hot core'' which has a gas density $n({\rm H}_2) \ga 10^7$
cm$^{-3}$ and a temperature of 150-300 K (see Kaufman et al. 1998 for
references). Whether this source is heated either externally (Blake et 
al. 1996, Wright, Plambeck \& Wilner 1996) or internally (Masson \& Mundy 
1988) is still a matter of great debate. In addition, Chandler \&
Wood (1997) suggested that there is an HII region associated with Radio
 Source ``I'', located to the northwest of the hot core ( 2$''$ away). 
This source has been proposed to be the heating source of the Orion hot core.
Nevertheless, as discussed in $\S1$, Kaufman et al. (1998) modeled the
physical conditions in the Orion hot core, and concluded that an internal
energy source is more likely, since the Radio Source ``I'' is not luminous
enough to externally heat the core up to the observed temperatures. Although 
high resolution molecular line and continuum studies have found evidence
of substructure within the core (Migenes et al. 1989; Blake et al. 1996), we 
apply our spherically symmetric model as a first approximation to this source.

The parameters of our best-fit model  are: a mass
accretion rate ${\dot M}= 1.1 \times 10^{-3}~M_{\odot}~{\rm yr}^{-1}$, a
B3 central star, an outer radius of the envelope $R_{\rm core}$ = 0.025
pc, and an opacity index $\beta = 1.6$ (see Table 2). The mass accretion
rate is determined mainly by the mm and submm data. The spectral type of
the star is constrained by the measured flux density at 30 $\mu$m
(Wynn-Williams et al.  1984), since an spectral type earlier than B3 
would have too much flux at this wavelength. Furthermore, the convolved image 
of the model at 1.3 mm has a FWHM of $\sim$ 3$''$, as observed by Blake 
et al. (1996; Figure 2, top panel). The total luminosity given by our model is 
$L_{\rm core}=2.5 \times 10^4~L_\odot$, being most of this due to 
the accretion. It is worth noting that this value exceeds the minimum 
luminosity required by the models of Kaufman et al. (1998)  to explain the 
observed high temperature in the Orion hot core.

\subsection{IRAS 23385+6053}

IRAS 23385+6053 has a quite broad data set available, spanning from
centimeter to near infrared wavelengths (see M98). No emission is 
detected at centimeter wavelengths  above  3-$\sigma$ level of $\sim$ 0.5 mJy.
At 3.4 mm, M98 detected a compact source with a deconvolved angular size of 
$4\rlap.''5 \times 3\rlap.''6$  (or 0.048 pc for an assumed distance of 4.9 
kpc). They estimated a bolometric luminosity for this source 
$L \sim 1.6 \times 10^4~ L_\odot$. They also detected a compact outflow in 
SiO and HCO$^+$ lines that is centered on the millimeter source.

In order to explain the lack of a detectable HII region associated with this 
source, M98 consider two possibilities: a heavily obscured B0 star with  
residual accretion, or a proto-star undergoing  very intense accretion with
${\dot M}=10^{-3} M_{\odot}$~yr$^{-1}$. For the latter case, using the 
relation found by Stahler et al. (1986), discussed in $\S 2.5$, they proposed 
a central proto-star with $M_*=39~M_{\odot}$ and $R_*=70~R_\odot$, and
assumed that the observed luminosity is dominated by the accretion luminosity.

Our best-fit (model I) to the observed spectrum is shown in the lower
right panel of Figure 3 and upper left panel of Figure 4. This model I
(whose parameters are given in Table 2) has a central B2 star, an
accretion rate of $10^{-3}~ M_{\odot}$~yr$^{-1}$, an external radius
$R_{\rm core}$ = 0.1 pc, and an opacity index $\beta$= 1.6. The physical
properties of this model are given in Table 3.  The lower left panel of
Figure 4 shows the intensity profile at 3.4 mm predicted by our model and
the observed intensity profiles along the major and minor axes of the 3.4
mm map given in Figure 1 of M98 (dot-dashed line). As can be seen in the
figure, the agreement is very good. The intensity profile of the model was
obtained by convolving (in two dimensions) the intensity distributions at
3.4 mm given by the model with a Gaussian beam of FWHM = $3\rlap.''8$
(corresponding to the size of the beam in the observations of M98). Note
that no fit was done to the observed intensity profile at 3.4 mm; only the
 observed flux densities (that are integrated values, with no
information on the spatial distribution) were fitted with a model spectrum
(upper panel of Fig. 4). It is remarkable that the predicted 3.4 mm model 
intensity profile, obtained from the parameter set that provide the best-fit to
the spectrum, is  in such a good agreement with the observed
intensity profile (lower panel of Fig. 4).

The  right  panels of Figure 4 shows the  predictions for an alternative  HMC
model, similar to  the one  proposed by M98 (model II), namely a core with a 
central proto-star with $M_*=39~M_{\odot}$, $R_*=70~R_\odot$,  and  
$L_{\rm core}=L_{\rm acc}$. This model has 
$ {\dot M} = 1.6 \times 10^{-3} M_{\odot}$ ~yr$^{-1}$, 
R$_{\rm core}=0.19$ pc  and $\beta =2$. Even though the observed spectrum can 
be fitted by this model, the predicted intensity profile at 3.4 mm is too 
broad and its peak intensity is too low to be detectable above the rms 
noise of the map (see lower right panel in Fig. 4). This difference 
is due the large value of $R_{\rm core}$, with respect to that of Model I. 
Therefore, this model appears to be in conflict with the observations. We 
conclude that a B2 star is most likely embedded in the IRAS 23385+6053 
core, as indicated by model I (Table 2 and 3 and left panel of Fig 4).

\subsection{Temperature Profiles of HMC models}

Figure 5 shows the temperature profiles for the models presented 
in Figure 3 (from $R_{\rm d}$ to $R_{\rm core}$). The different sources are 
labeled in each panel. These temperature profiles are discontinuous at the 
photospheric radius, $R_{\rm ph}$, because the dilution 
factor, $W(R_{\rm ph}) = 1/2$, and, in our approximation,  eqn. (5) then 
gives $T < T_{\rm ph}$ at this radius. Instead, there should be a smooth 
transition between the optically thin and the optically thick regimes and 
hence a smooth temperature profile. With our simplified treatment for the 
temperature profile, the difference between the total luminosity of the 
spectra of model and the input luminosity,  $L_{\rm core}$, is less than 
15 - 20 $\%$. It can be seen  that the temperature distributions in the 
optically thick region have a steeper dependence with radius than in the 
optically thin region, as discussed  by Adams $\&$ Shu (1985) and  Kenyon 
et al. (1993). Finally, for all cases, at the external radius, $R_{\rm core}$, 
the temperatures are $T > 30 K$.

\section{DISCUSSION}
\subsection{General Discussion}

The model studied here, of an envelope infalling onto
a central massive star, reproduces very well the currently available 
observational data for the sources selected in our sample.  
We have explored two possible density distributions for the structure of
the infalling envelope and concluded that a SLS collapse density distribution 
provides a better agreement with the available observational data, than a SIS 
density distribution. In order to fit the observations of these sources,
our model requires massive, early B-type central stars (10-16 $M_\odot$), 
with high values of the mass accretion rate 
($\sim 10^{-3}~M_\odot$ yr$^{-1}$), and core radii  $\la 0.1$~pc.
The values of the mass accretion rate obtained  exceed by a large factor 
the ``critical'' value required to quench the development of a detectable
UCHII region, which for an early B-type central star (the proposed central
stars), is $\la 10^{-7}~M_\odot$ yr$^{-1}$. Therefore, the model lacks a
detectable HII region. Furthermore, very small flux densities from dust
thermal emission are predicted at centimeters wavelengths.

The values that we have obtained for the radius of the cores are roughly 
consistent with those inferred from the observations. However, a detailed 
comparison is not yet possible, since most of the observations give 
only upper limits for the source sizes. Only in the case of the 3.4 mm map 
of IRAS 23385+605 (M98), was it  possible  to compare the intensity
distribution predicted by the model  with that observed, and we found
 a remarkably good agreement with the observed total flux, source size and 
peak intensity. Additional maps of good S/N and high angular resolution are 
required for a larger number of sources and wavelengths. In this manner, 
continuum intensity profiles may be able to discriminate between possible 
models. Similarly, high angular resolution FIR images are needed. Our model 
predicts a peak flux density $>10^3$ Jy at $\sim 100~ \mu m$. Thus, infrared 
fluxes around this wavelength, obtained with higher angular resolution than 
currently available to avoid contamination with nearby objects, would be 
specially relevant in order to further constrain the parameters of the 
modeled sources.

The intrinsic infrared colors  [100-60], [60-25]  and [25-12]  predicted by
our models (see Table 4) are very different from those of observed UCHII 
regions, [100-60]=0.26, [60-25]=0.87, and [25-12]=0.91 (Wood \& Churchwell 
1989). In particular, a color criterion [60-25] $>$ 1.3 could be use to find 
isolated HMCs, when their fluxes are not contaminated by nearby sources.

Furthermore, for all the sources, the accretion luminosity clearly exceed, 
the stellar luminosity (see Table 3). This result points out the 
relevance of the energy released in the accretion process at this early 
evolution stage, a basic ingredient that should not be ignored when modeling
this kind of object.
  
\subsection{Radiation Pressure Onto  Dust  Grains}

The radiation pressure onto dust grains imposes an upper limit to the
amount of mass that can accrete onto a massive star. It is therefore important
to assess whether or not  this effect can halt the collapse of the envelopes 
that we  consider in our models. Kahn (1974) found that the
radiation pressure of the diffuse infrared radiation field inside a dusty
envelope around a massive proto-star decelerates the flow allowing a
maximum luminosity to mass ratio $L/M_* = 5436~L_{\odot}/ M_{\odot}$,
 although this number depends on the properties of the dust grains.  Wolfire \&
Cassinelli (1987; hereafter WC87) revised the calculations of the maximum
mass that an accreting proto-star can accumulate taking into account updated
properties of the interstellar grains. Jijina \& Adams (1996) reexamined
this problem including the effect of angular momentum. In this section,
 we ensure that our  models of hot cores  fulfill the general constraint that
 guarantee that inflow will proceed towards the central proto-star.

At the outer boundary of the inflow, the outward radiative acceleration
must be less than the inward gravitational acceleration of
the gas. This condition can be written as (WC87, eqn. [10])

\begin{equation}
\Gamma = {\kappa^{\rm pr}_{\rm H} L_{\rm core} \over 4 \pi c 
G M(R_{\rm core})} < 1, 
\end{equation}
where $\kappa^{\rm pr}_{\rm H}$ is the flux mean radiation 
pressure coefficient, $c$ is the speed of the light, $M(R_{\rm core})$ is 
the total mass inside
 $R_{\rm core}$ (star plus core mass), and $L_{\rm core} = L_{*}
 + L_{\rm acc}$ is the total luminosity available to heat and push the
 core material.  Our models have $L_{\rm core} \la  6 \times 10^4~
L_\odot$, typical core masses $M(R_{\rm core}) \sim 130~ M_\odot$, and we
approximate the flux mean opacity with the dust Planck mean opacity
evaluated at the photospheric temperature.  The photospheric temperatures
we find are $T_{\rm ph} \sim 100~K$, therefore, $\kappa_P \sim 2~ \op$. For 
this range of values 
\begin{equation}
\Gamma = 0.07 \ \left( {\kappa_P \over 2 \ \op } \right) 
\left( {L_{\rm core} \over 6 \times 10^4~L_\odot} \right)
\left( {M(R_{\rm core}) \over 130~M_\odot} \right)^{-1} < 1, 
\end{equation}
and the fluid at the outer boundary can flow inward.

On the other side, at the dust destruction front, $ R_{\rm d}$, where 
the direct stellar radiation is absorbed, the momentum rate of the flow 
must be greater or equal to the pressure of the stellar radiation field 
\begin{equation}
\dot M \ge {L \over u_{\rm d} c}, 
\end{equation}
where $u_{\rm d}$ is the velocity of the flow at the dust destruction front,
$ R_{\rm d}$. This condition is satisfied for our model ($M_{*} \sim
 10-16~M_\odot$, $\dot M \sim 6 \times 10^{-4} - 10^{-3}~M_\odot~
 {\rm yr}^{-1}$) as can be seen in Fig. 5 of WC87.
  Their figure also shows
 that our cores have less than the Eddington luminosity, $L_{\rm Edd}$, where
 the accretion is halted by radiation pressure on free electrons with an
opacity given by Thompson scattering. 

Considering an optically thick dusty cocoon plus an optically thin
envelope, and assuming an analytic form for the Rosseland mean opacity
$\chi_{\rm R} = \chi_{\rm R0} (T/T_s)^2$, Kahn (1974) obtained the maximum
luminosity to mass ratio $L/M_* = 5436~L_{\odot}/ M_{\odot}$,  mentioned above.
 His eqn. [65] implies that this ratio is inversely proportional both to
 the cube of the dust sublimation temperature, $T_{\rm sub}^{3}$, and
 to $\chi_{R0}$, the Rosseland mean opacity for radiation at a
 color temperature $T_{\rm s} = 22,000$ K.  Kahn used $T_{\rm sub} = 3675$ K
 and $\chi_{R0} = 600~\op$. We find that the sublimation temperature
 for the grain mixture in our models (see $\S 2.4$) is $T_{\rm sub} \sim
 1200$ K.  Also, the value of $\chi_{R0}$ must be increased by a factor of
 $\sim 8 $ to agree with our value for the Rosseland mean opacity at 
 $T \sim 1000$ K.  These modifications, due to the currently accepted
 properties
 of the standard grain mixture used in this work, imply that the maximum 
luminosity to mass ratio derived by Kahn would currently be
$L/M_* \sim 20,000~L_{\odot}/ M_{\odot}$.
This result contrasts with the detailed modeling of the dust
 destruction process carried out by WC87.
 They found that very massive stars can only be formed with modifications of
 the standard MRN grain mixture for
accretion onto massive stars with  $M_*$ = 60, 100 and 200 $M_\odot$,
 in  which the luminosity to mass ratios are larger than $L/M_* = 8974
 ~ L_{\odot}/ M_{\odot}$.
 On the other hand, Jijina $\&$ Adams (1996) found that the maximum luminosity
to mass ratio is less restrictive for a rotating collapse than for
the spherical infall.  In any case, all our models have luminosity to
 mass ratios $L_{\rm core}/M_* < 4300 ~L_{\odot}/ M_{\odot}$, which
 implies that accretion can continue
 onto the central star. 

We therefore consider that radiation pressure onto dust grains will not
stop the accretion flow onto the central proto-star for  the modeled HMCs 
discussed in this work. For simplicity, we have not considered  any
deviations of the flow from the dynamics of a logatropic collapsing core. 

\subsection{End of the Accretion Phase}

The luminosity to mass ratios of the modeled HMCs discussed in this work are 
high but still remain below the critical ratios for radiation pressure onto 
dust grains needed to halt  the accretion (see $\S 5.2$). Nevertheless, this 
ratio, given by
\begin{equation}
{L_{\rm core} \over M_*} = {L_* \over M_*} + {G \dot M_* \over R_*},
\end{equation}
will increase with time both because the luminosity of the central star 
increases with time, and because the mass accretion rate increases as
 $\dot M \propto t^3$ (eqn. [1]) in the logatropic model. Therefore, 
radiation pressure could eventually halt the collapse of matter onto the 
central star. This is only a transient process since, as the accretion is 
shut off, so is the accretion luminosity, an important contributor to the 
total luminosity.  Thus, one is tempted to speculate that, if a stellar wind 
can turn on after the flow is impulsively reversed, the wind could help clear 
out the accreting flow. An HII region can then be produced, although in this 
scenario its evolution would be governed by the reversal of the accretion 
flow, a complex problem worth studying in detail.
 
If this process occurs on a  short time scale compared to the accretion time 
scale, one can make a simple estimate of the final masses of the central 
stars inside the HMCs when the accretion flow is halted by radiation 
pressure. We assume a main sequence mass-luminosity relation, 
$L_* \propto M_*^{3}$, and mass-radius relation, 
$R_* \propto M_*^{0.73}$ (e.g. Bowers \& Deeming 1984), even though,
as discussed in $\S$ 2.5, it is not clear if these relations hold for 
massive stars formed under  intense accretion rates which increase steeply 
with time, considered in this work. Figure 6 shows the luminosity to mass 
ratio as function of the dimensionless time $t/t_{\rm age}$, where 
$t_{\rm age}$ is the present age of the HMC (solid line). 
We have assumed a $12.6~M_\odot$ central star and a mass accretion rate 
$\dot M = 1 \times 10^{-3}~M_\odot~{\rm yr}^{-1}$, at the dimensionless time 
$t/t_{\rm age} = 1$. These values correspond to the model for IRAS 23385+6053 
which has $t_{\rm age}= 5 \times 10^4$ yr (see Tables 2 and 3). 
The dot-dashed line is the stellar contribution to the luminosity to mass 
ratio and the dashed line is the accretion contribution to this ratio. The 
triangles show the mass of the central star ($M_* \propto t^4$, see eqn. 
[1]) labeled on the right hand vertical axis. Independently of the exact 
critical value for the luminosity to mass ratio for radiation pressure to 
halt the accretion flow (see discussion in $\S 5.2$), this figure shows that 
the evolution of this ratio is very fast so one would expect that in less 
than $10^5$ yr  the accretion of mass onto the central star of the HMC 
should stop and an ultracompact HII region should develop inside these cores.

Finally, if all OB stars are formed within HMCs under high accretion rates, 
one can estimate the expected number of HMCs in this phase. Taking the rate 
of formation of O stars in the Galaxy $\dot N \sim 1/300~{\rm yr}^{-1}$ 
(Wood \& Churchwell 1989), and the time in the HMC phase  when the central 
star is a late B star, $\Delta t \sim 10^5~{\rm yr}$, one should expect to 
find $\sim 300$ HMCs at the present time. On the other hand, the period of 
time that a core contains an early O star is very small, 
$ \Delta t \lesssim 2 \times 10^4~{\rm yr}$, so less than 60 HMCs 
with very luminous massive stars inside  should be expected. The number of 
expected HMCs can increase by a factor of 10 if a higher star formation 
rate of O stars,  $\sim 4 \times 10^{-2}~{\rm yr}^{-1}$ (G\"usten \& Mezger 
1983), is used.

\section{CONCLUSIONS}

Our main results can be summarized as follows:

\begin{enumerate}

\item The observed spectra of several HMCs (G34.24 + 0.13MM, 
W3(H$_2$0), the Orion hot core and IRAS 23385+6053) can be reproduced by  
massive envelopes accreting onto  young massive central stars. In order to 
fit the available data, we require early B-type stars with high mass 
accretion rates, ${\dot M} \gtrsim 6\times10^{-4} M_{\odot}$ yr$^{-1}$, and 
time-scales $t_{\rm age} \lesssim 10^{5}$ yr. The main heating agent 
is the accretion luminosity (L$_{\rm acc} > L_*$; see Table 3).

\item The spectra are strongly dependent on the assumed envelope density
profile. Isothermal collapsing envelopes have less mass in the envelope for a 
given mass accretion rate, with respect to the logatropic collapsing 
envelopes. Therefore, they  require very high accretion rates ($> 10^{-3}
 M_{\odot}$ yr$^{-1}$) in order to reproduce the observed millimeter 
emission. Nevertheless, high accretion rate SIS models predict strong 
mid-infrared emission which is, in general, not observed. 

\item Models of the structure of massive stars formed under  intense mass 
accretion rates that increase with time, discussed in this work, are 
necessary to determine the characteristics (e.g. radius, intrinsic 
luminosity) of this type of object. In particular, we have used central stars 
with radii larger than the ZAMS radii in order to avoid excess fluxes
at mid infrared wavelengths with respect to those observed.

\item High angular resolution mid and far IR observations (which avoid 
contamination by  nearby massive stars) are crucial to constrain the total 
luminosity of the HMCs. 

\item The critical mass accretion rate required to quench the development of 
an UCHII region  is very small compared to the accretion rates required by 
all the HMC models studied here. Therefore, at this stage the HMCs have no 
detectable HII region. 

\item We speculate that  when a critical luminosity to mass ratio is 
achieved, the infall will be halted by a powerful stellar wind and a 
detectable UCHII region can appear.
\end{enumerate}

\acknowledgments
 
We are greatly indebted with Javier Ballesteros, Mar\'\i a
Eugenia Contreras, and Rosa Izela D\'\i az-Miller
with whom we originally discussed the work presented here.
We also thank Guillem Anglada, Riccardo Cesaroni, Guido Garay, William 
Henney, Stan Kurtz, Sergio Molinari, Luis F. Rodr\'\i guez, Leonardo Testi,
 Malcom Walmsley and Alan Watson for useful discussions and 
helpful suggestions. We thank Leonardo Testi for providing us the
original data for IRAS 23385+6053. Thanks too to Robert 
Estalella for providing a program to convolve  our results with the observing 
beam. M. O., S. L. and P. D. acknowledge support from DGAPA/UNAM and 
CONACYT.  S. L. also acknowledges support from the  Simon  Guggenheim 
Memorial Foundation. M. O. also acknowledges partial support from the 
Programa de  Cooperaci\'on Cient\'\i fica con Iberoam\'erica (Spain).

\clearpage
\begin{deluxetable}{cccccccc}
\tablecaption{\sc Physical Properties of a Grid of SLS Hot Core Models
\label{tbl-1a} }
\scriptsize
\tablewidth{0pt}
\tablenum{1a}
\tablehead{
\colhead{ Model }
&\colhead{L$_{\rm core}$}
&\colhead{R$_{\rm d}$} 
&\colhead{R$_{\rm ph}$}
&\colhead{n($R_{\rm core}$)} 
&\colhead{T$_{\rm ph}$}
&\colhead{r$_{\rm ew}$}
&\colhead{M$_{\rm core}$}\nl
\colhead{}
&\colhead{($10^4$L$_{\odot}$)}
&\colhead{(AU)}
&\colhead{($10^3$AU)}
&\colhead{($10^5$cm$^{-3}$)}
&\colhead{(K)}
&\colhead{($10^3$AU)}
&\colhead{(M$_{\odot}$)}
}	       
\startdata
B0 star; ~ R$_{\rm core}$=0.1 pc \hfill &&&&&&  \nl
${\dot M} $ ($M_{\odot}~{\rm yr^{-1}}$)&& & & & \nl
\hline
1.5$\times 10^{-3}$ & 16 & 118   & 11.0   & 14 & 74 &21  &495\nl
4.4$\times 10^{-4}$ &6.4 &  47   & 3.2    & 2.6& 110&47  & 86\nl
8.1$\times 10^{-5}$ &3.2 &  18   & 0.4    & 0.2& 260&144 &  8\nl
\hline
& \nl
R$_{\rm core}$=0.1 pc; ${\dot M}=4.4\times10^{-4}$  $M_{\odot}
~{\rm yr}^{-1}$& & &&&& \nl
Spectral type&&  & &&&\nl
\hline
O7                  &15.2 & 57   & 2.3 & 1.1  & 162  &95 & 36\nl
B0                  & 6.4 & 47   & 3.2 & 2.6  & 110  &47 & 86\nl
B3                  & 4.3 & 45   & 4.2 & 4.3  &  87  &30 &146\nl
\hline
& \nl
B0 star; ${\dot M}=4.4\times10^{-4}$
$M_{\odot}~{\rm yr}^{-1}$ &&&&&&\nl
$R_{\rm core}$ (pc)& &&&  &&\nl
\hline
0.15                & 6.4 & 47   & 3.9  & 1.8  & 100 &47 &205\nl
0.10                & 6.4 & 47   & 3.2  & 2.6  & 110 &47 & 86\nl
0.05                & 6.4 & 47   & 2.3  & 4.3  & 130 &47 & 19\nl
\enddata
\end{deluxetable}


\begin{deluxetable}{cccccccc}
\tablecaption{\sc Physical Properties of a Grid of SIS Hot Core Models
\label{tbl-1b} }
\scriptsize
\tablewidth{0pt}
\tablenum{1b}
\tablehead{
\colhead{ Model }
&\colhead{L$_{\rm core}$}
&\colhead{R$_{\rm d}$} 
&\colhead{R$_{\rm ph}$}
&\colhead{n($R_{\rm core}$)} 
&\colhead{T$_{\rm ph}$}
&\colhead{r$_{\rm ew}$}
&\colhead{M$_{\rm core}$}\nl
\colhead{}
&\colhead{($10^4$L$_{\odot}$)}
&\colhead{(AU)}
&\colhead{($10^3$AU)}
&\colhead{($10^5$cm$^{-3}$)}
&\colhead{(K)}
&\colhead{($10^3$AU)}
&\colhead{(M$_{\odot}$)}
}	          
\startdata
B0 star; ~ R$_{\rm core}$=0.1 pc \hfill &&&&&& \nl
${\dot M} $ ($M_{\odot}~yr^{-1}$)&& & & & \nl
\hline
1.5$\times 10^{-3}$ & 16 & 115   & 6.5  &2.3 & 98&  4.1 &141\nl
4.4$\times 10^{-4}$ &6.4 &  47   & 2.1  &1.0 &136&  9.4 & 47\nl
8.1$\times 10^{-5}$ &3.2 &  18   & 0.3  &0.1 &284&  29  &  4\nl
\hline
& \nl
R$_{\rm core}$=0.1 pc; ${\dot M}=4.4\times10^{-4}$  $M_{\odot}
~{\rm yr}^{-1}$& & &&&&\nl
Spectral type&&  & &&&\nl
\hline
O7                  &15.2 & 56   & 1.6  &0.9 & 192& 19.0 & 20\nl
B0                  & 6.4 & 47   & 2.1  &1.0 & 136&  9.4 & 47\nl
B3                  & 4.3 & 44   & 2.5  &1.0 & 114&  6.0 & 57\nl
\hline
& \nl
B0 star; ${\dot M}=4.4\times10^{-4}$
$M_{\odot}~{\rm yr}^{-1}$ &&&&&&\nl
$R_{\rm core}$ (pc)& &&&  &&\nl
\hline
0.15                & 6.4 & 47   & 2.3  &0.4 & 130& 9.3  &84\nl
0.10                & 6.4 & 47   & 2.1  &1.0 & 136& 9.4  &47\nl
0.05                & 6.4 & 47   & 1.6  &4.0 & 157& 9.4  &11 \nl
\enddata
\end{deluxetable}

\clearpage

\begin{deluxetable}{cccccccc}
\tablecaption{\sc Model Parameters Determined from Spectral Fits to 
Individual Sources  \label{tbl-2} }
\tablewidth{0pt}
\tablenum{2}
\tablehead{
\colhead{Source}
&\colhead{$M_{*}^a$}
&\colhead{$L_{*}^b$}
&\colhead{${\dot M}$}
&\colhead{$R_{\rm core}$} 
&\colhead{$\beta$}
&\colhead{$D^c$}\nl
&\colhead{(M$_{\odot}$)} 
&\colhead{(L$_{\odot}$)}
&\colhead{(M$_\odot$ yr$^{-1}$)}
&\colhead{(pc)}
&\colhead{}
&\colhead{(kpc)}
}

\startdata
G34.24+0.13MM  & 10.0 &1.0$\times 10^{3}$&6.5$\times 10^{-4}$ & 0.07 & 2.0
& 3.7 \nl
& \nl 
W3(H$_2$O)     & 15.8 &2.5$\times 10^{4}$&1.2$\times 10^{-3}$ & 0.05 & 1.6
& 2.2 \nl
& \nl
Orion Hot Core & 10.0 &1.0$\times 10^{3}$&1.1$\times 10^{-3}$ & 0.02 & 1.6
& 0.5  \nl
& \nl
IRAS 23385+6053& 12.6 &2.8$\times 10^{3}$&1.0$\times 10^{-3}$ & 0.10 & 1.6
& 4.9  \nl
& \nl
\enddata
\tablenotetext{a}{ $R_* = 10^{12}$ cm is assumed.}
\tablenotetext{b}{ Taken from Thompson (1984), for a given $M_*$.}
\tablenotetext{c}{ The adopted distances are taken from Kuchar $\&$ Bania 
1994, Humphreys 1978, Genzel et al. 1981, and Molinari et al. 1998.}
\end{deluxetable}


\begin{deluxetable}{lcccccccccc}
\tablecaption{\sc Physical Properties of Best-Fit Models to Individual HMCs
\label{tbl-3} }
\tablewidth{0pt}
\tablenum{3}
\scriptsize
\tablehead{
\colhead{Source}
&\colhead{$R_{\rm d}$} 
&\colhead{$R_{\rm ph}$}
&\colhead{$T_{\rm ph}$}
&\colhead{$n(R_{\rm core})$}
&\colhead{$L_{\rm core}$}
&\colhead{$L_{\rm acc}/L_{\rm core}$}
&\colhead{$M_{\rm core}$}
&\colhead{$N(H_2)$}
&\colhead{$t_{\rm age}$}
&\colhead{$r_{\rm ew}$}\nl
&\colhead{(AU)}
&\colhead{($10^3$ AU)} 
&\colhead{(K)} 
&\colhead{({$10^6$ cm$^{-3}$})}
&\colhead{($10^4$ L$_{\odot}$)}
&\colhead{$(\%)$}
&\colhead{(M$_\odot$)} 
&\colhead{($10^{24}$ cm$^{-2}$)}
&\colhead{($10^4$ yr)}
&\colhead{($10^3$ AU)}
}
\startdata
G34.24+0.13MM   &35&3.8&72&1.0&1.6&94&131&34&6.1&20 \nl
& \nl
W3(H$_2$O)      &74&5.3&87&2.0&6.7&62&83&48&5.3&24 \nl
& \nl
Orion hot core  &52&3.2&88&5.0&2.5&96&27&55&3.7&16 \nl
& \nl
IRAS 23385+6053 &54&7.0&62&1.1&3.1&91&382&47&5.0&21 

\enddata
\end{deluxetable}


\begin{deluxetable}{lccc}
\tablecaption{\sc Infrared Colors of the Best-Fit Models
\label{tbl-4} }
\tablewidth{0pt}
\tablenum{4}
\tablehead{
\colhead{Source}
&\colhead{[100 - 60]}
&\colhead{[60  - 25]}
&\colhead{[25  - 12]} 
}

\startdata
G34.24+0.13MM   & 0.04 & 2.32 & 4.60   \nl
& \nl 
W3(H$_2$O)      &-0.18 & 1.36 & 3.58   \nl
& \nl
Orion hot core  &-0.16 & 1.27 & 3.43   \nl
& \nl
IRAS 23385+6053 & 0.20 & 2.86 & 5.84   \nl
& \nl
\enddata
\end{deluxetable}

\clearpage

\begin{figure}

\figcaption[]{\label{fig1}(a) Spectra for different
sets of the SLS collapse density distribution HMC models. The top panel shows 
the trends of the spectra obtained for three different values of the mass 
accretion rate, for a B0 central star and an external radius 
$R_{\rm core}=0.1$ pc. The middle panel shows models with  different spectral 
types of the central star, for a given external radius $R_{\rm core}= 0.1$ pc 
and accretion rate of $ {\dot M} = 4.4\times 10^{-4}~M_{\odot}$ yr$^{-1}$. 
The bottom panel shows models with  different external radii of the envelope, 
for a central B0 star and an accretion rate 
$ {\dot M} = 4.4\times 10^{-4}~M_{\odot}$ yr$^{-1}$. (b) Same as 1(a) but 
for the SIS collapse density distribution HMC models. In all  cases, a 
distance of 4.9 kpc were adopted. }

\end{figure}


\begin{figure}
\figcaption[]{ \label{fig2}
Observed flux densities of the source G34.24+0.13MM (Hunter et al. 1998; Goss 
1999, personal communication). Three modeled spectra with $R_{\rm core}$ = 
0.07 pc and a central B3 star (10 $M_{\odot}$) are also shown. The dotted 
line corresponds to the collapse density distribution of the SIS, with  
${\dot M} = 1.0 \times 10^{-3}~M_{\odot}~{\rm yr}^{-1}$, and the dot-dashed 
line corresponds to the collapse density distribution of the SLS, with 
${\dot M} = 6.5 \times 10^{-4}~M_{\odot}~{\rm yr}^{-1}$, when ZAMS radii are 
adopted. The solid line shows the spectrum of the SLS model but with a star 
that has a large radius, $R_* = 10^{12}$ cm.}
\end{figure}


\begin{figure}
\figcaption[]{ \label{fig3}
Observed flux densities for the sources G34.24+0.13MM (Hunter et al. 1998; 
Goss 1999, personal communication), W3(H$_2$O) (Wyrowski et al. 1997;
Wilner et al. 1995), Orion hot core (Wright et al. 1992; Murata et al. 1991;
Masson et al. 1985; Mezger et al. 1990; The fluxes and errorbars at 
30 $\mu$m and 1.3 mm was estimated from the maps of  Wynn-Williams et al. 
1984 and Blake et al. 1996, respectively), and IRAS 23385+6053 
(Molinari et al. 1998; IRAS-PSC2), and model spectra. The parameters of each 
model are listed in Table 2.}
\end{figure}


\begin{figure}
\figcaption[]{ \label{fig4}
Observed flux densities (Molinari et al. 1998; IRAS-PSC2) and 
two models for the source IRAS 23385+6053. The upper left panel shows the 
spectrum for a model I, with B2 central star, $\dot M = 10^{-3}~ M_\odot
~ {\rm yr}^{-1}$, $R_{\rm core} = 0.1$ pc, and $\beta=1.6$. 
The lower left panel shows the intensity profile of the 3.4 mm emission
given by the model (solid line) and the observed intensity profiles along the 
major and minor axes of the 3.4 mm map given  in Fig. 1 of M98 (dot-dashed 
lines). The upper right and lower right panels correspond to  model II with 
$\dot M = 1.6 \times 10^{-3}~M_\odot {\rm yr}^{-1}$, $R_{\rm core} = 0.19$ 
pc, and $\beta =2$, and a central proto-star with $M_* = 39~M_\odot$ and 
$R_* = 70~R_\odot$ (for this model, $L_{\rm core} = L_{\rm acc}$ and $L_*$
is neglected)}
\end{figure}


\begin{figure}
\figcaption[]{\label{fig5}
Temperature profiles in the range $R_{\rm d} < r < R_{\rm core}$, for the 
models shown in Figure 3. The different sources are labeled in each panel. 
The parameters of each model are given in Table 2. The dust destruction
radius, $R_{\rm d}$, the photospheric radius, $R_{\rm ph}$, and the 
photospheric temperature, $T_{\rm ph}$, for each source are given in Table 3.}

\end{figure}


\begin{figure}
\figcaption[]{ \label{fig6}
Luminosity to mass ratio as function of the dimensionless time 
$t/t_{\rm age}$ (solid line). The dot-dashed line is the stellar 
contribution to this ratio and the dashed line is the accretion 
contribution. The triangles show the mass of the central star, 
labeled  on the right hand vertical axis.}
\end{figure}

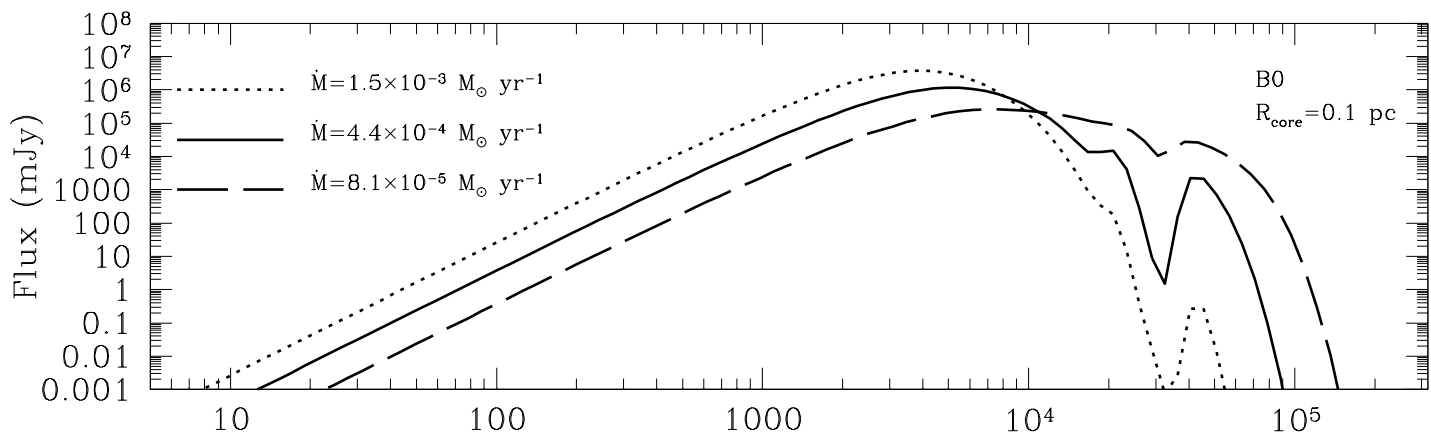
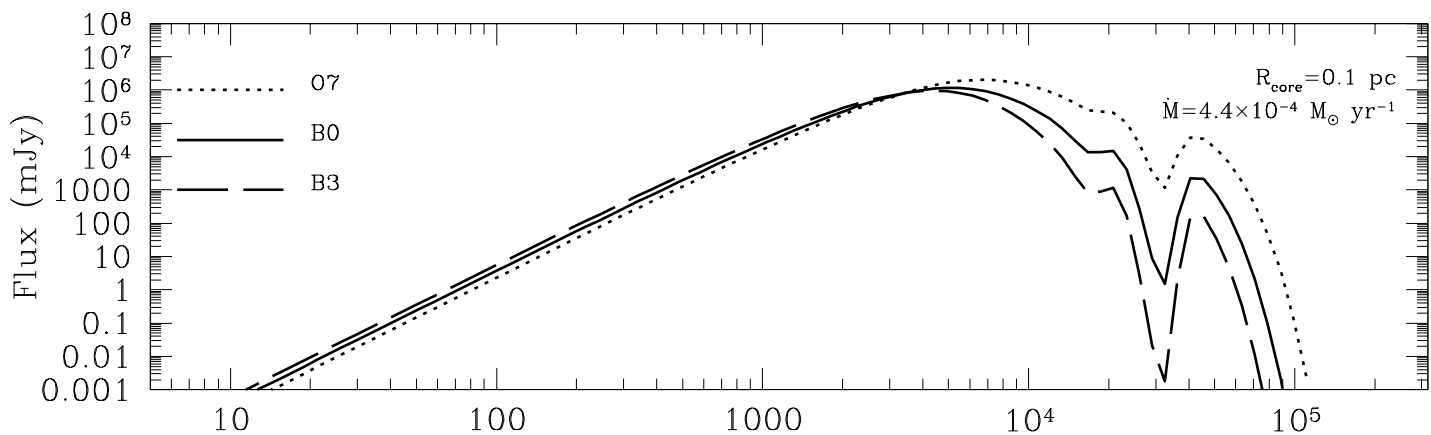
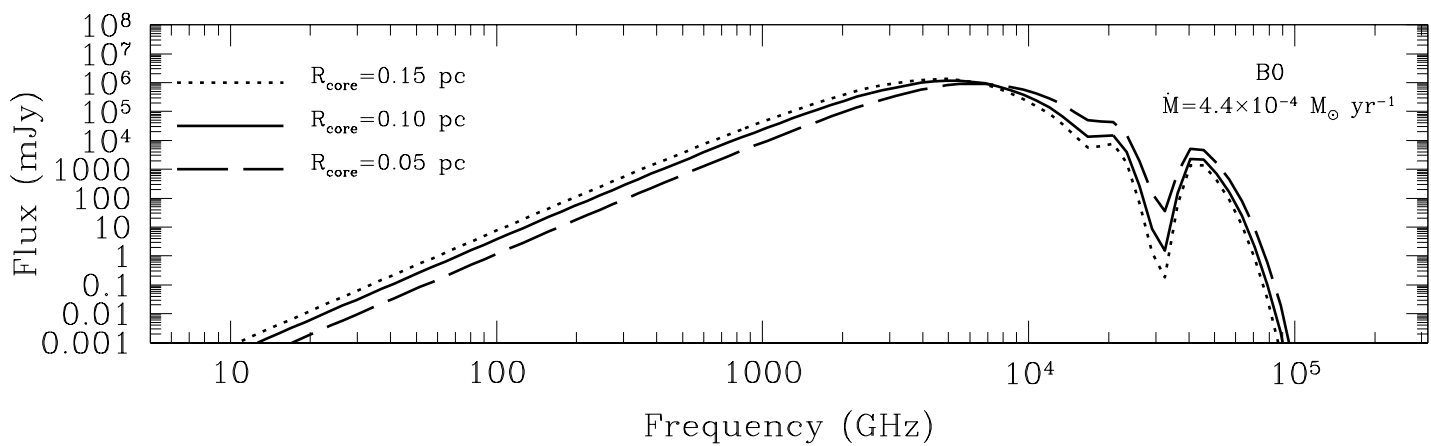
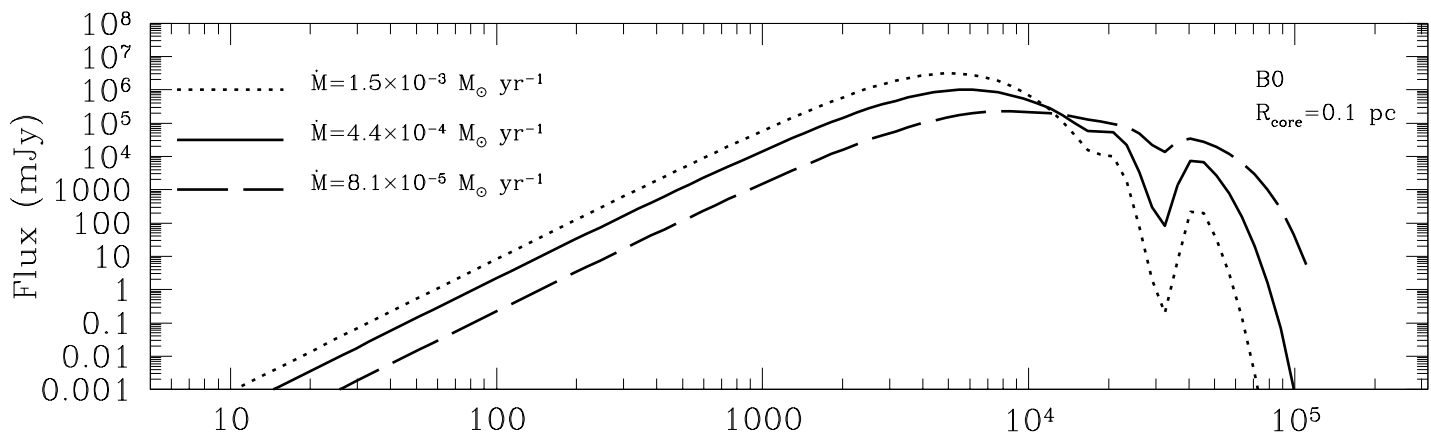
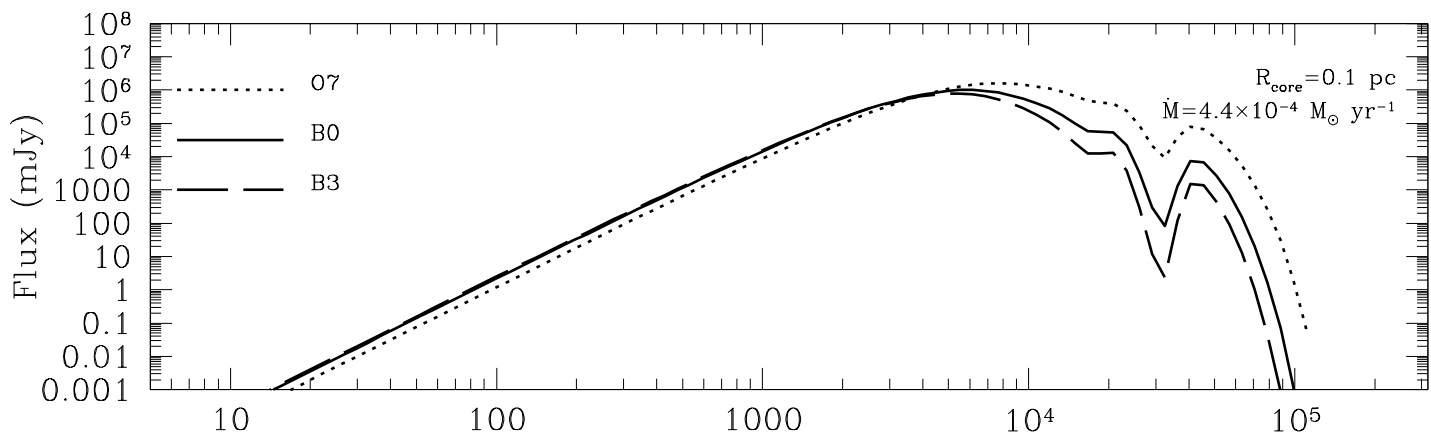
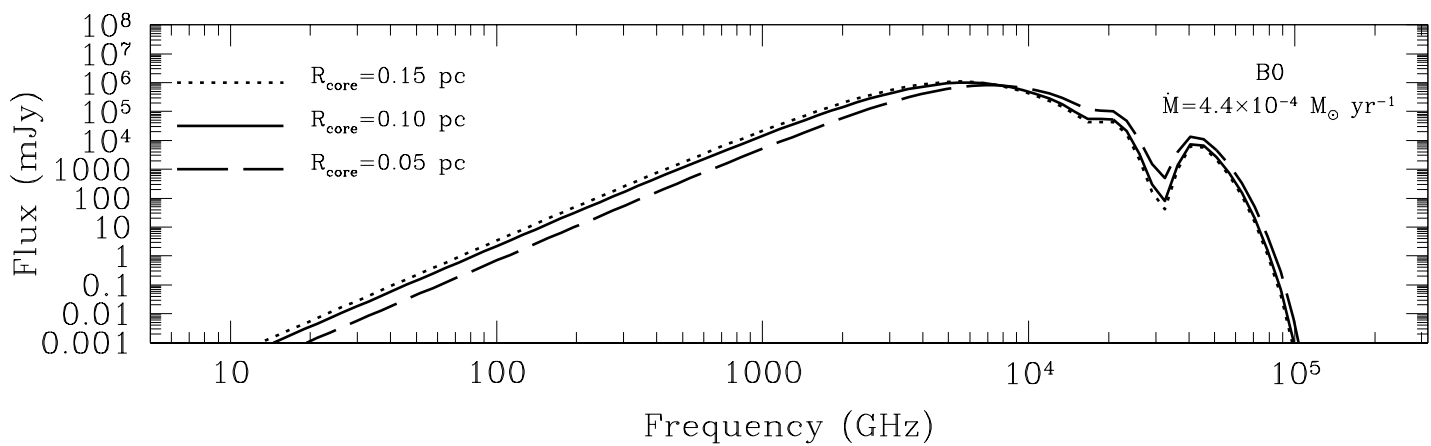
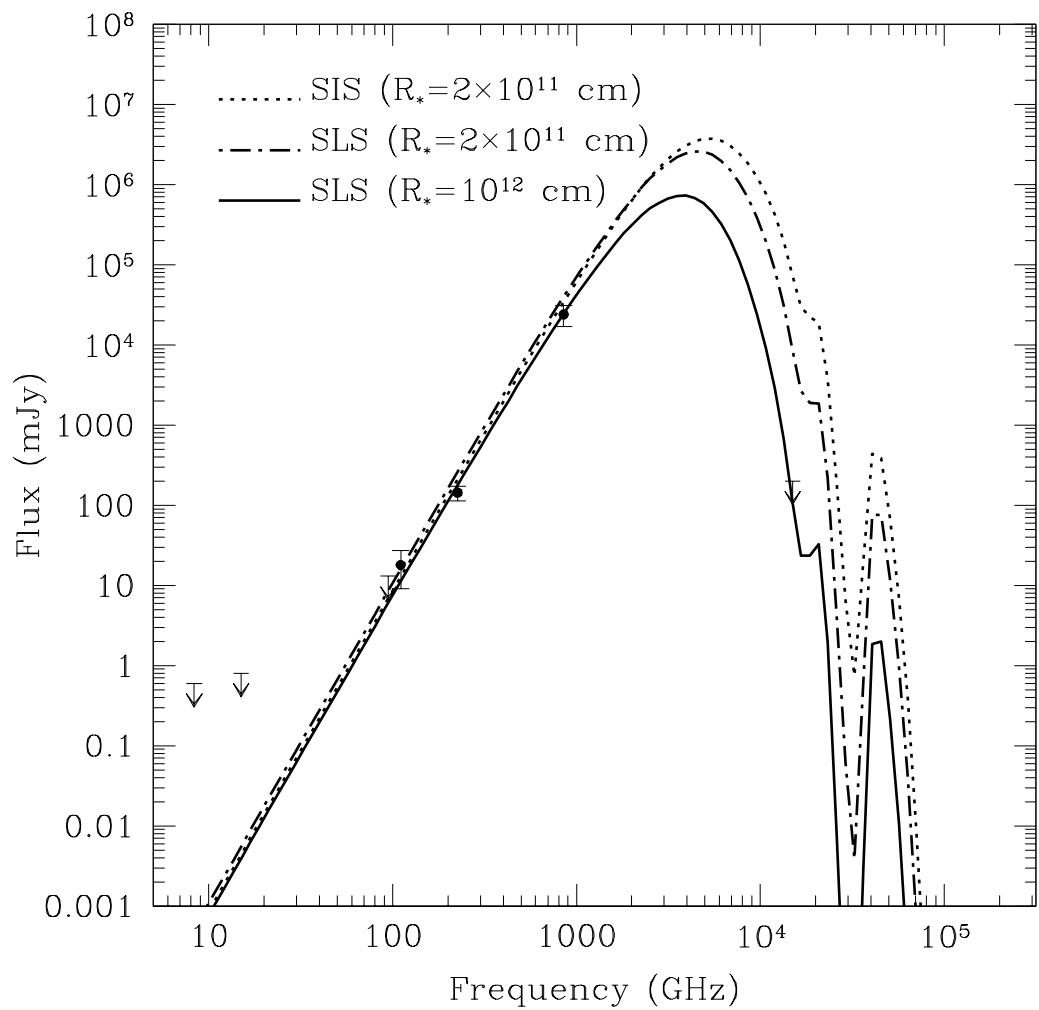
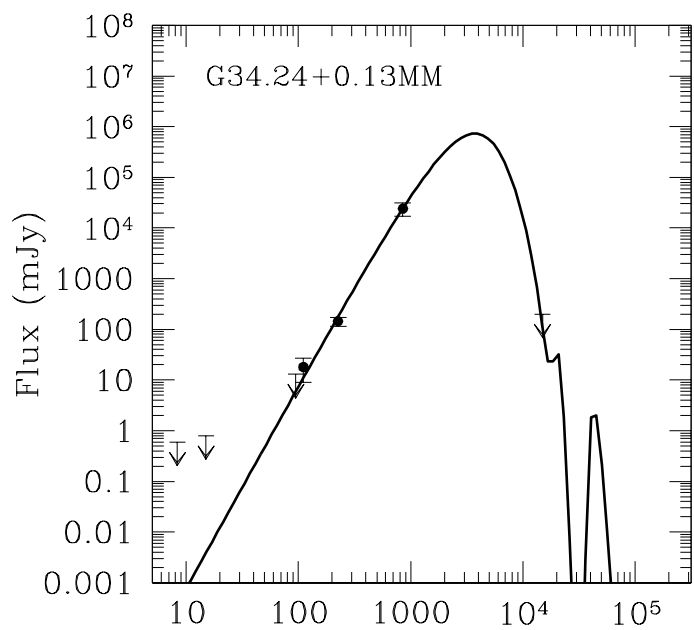
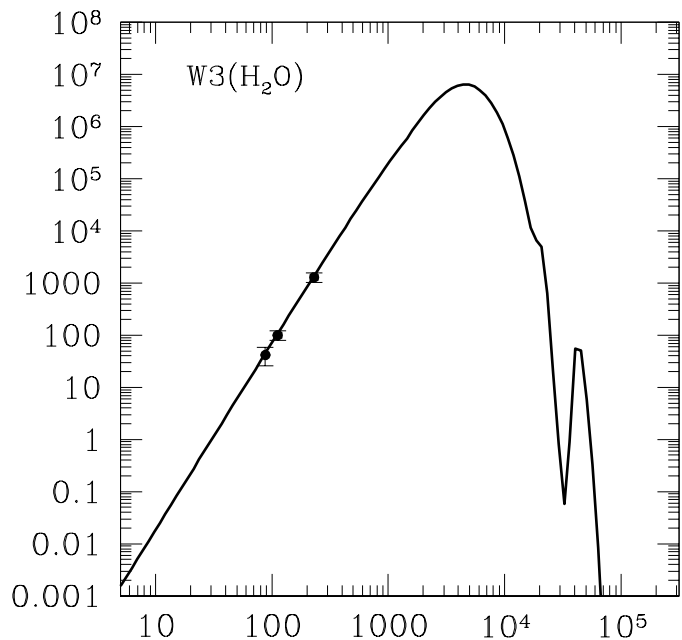
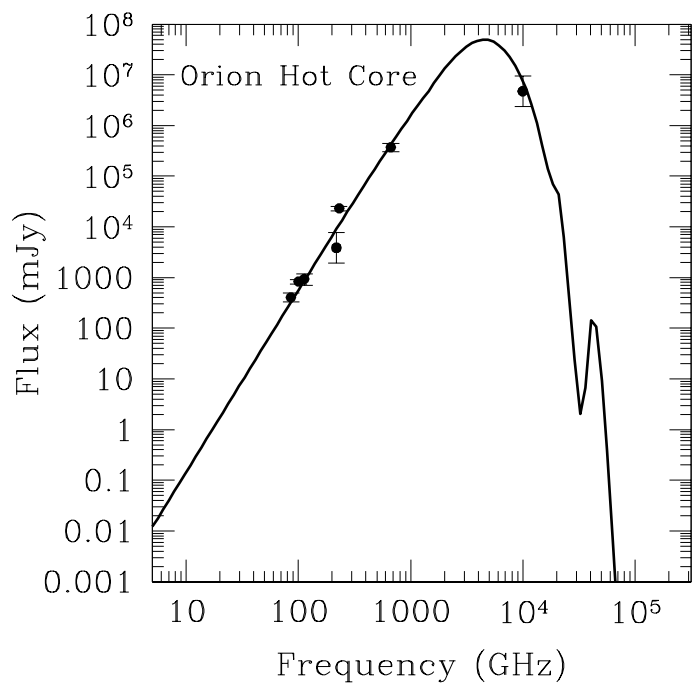
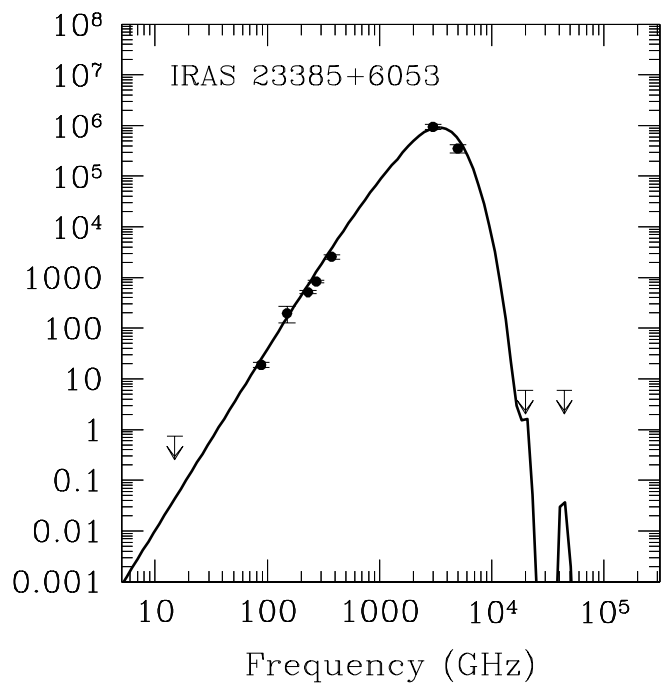
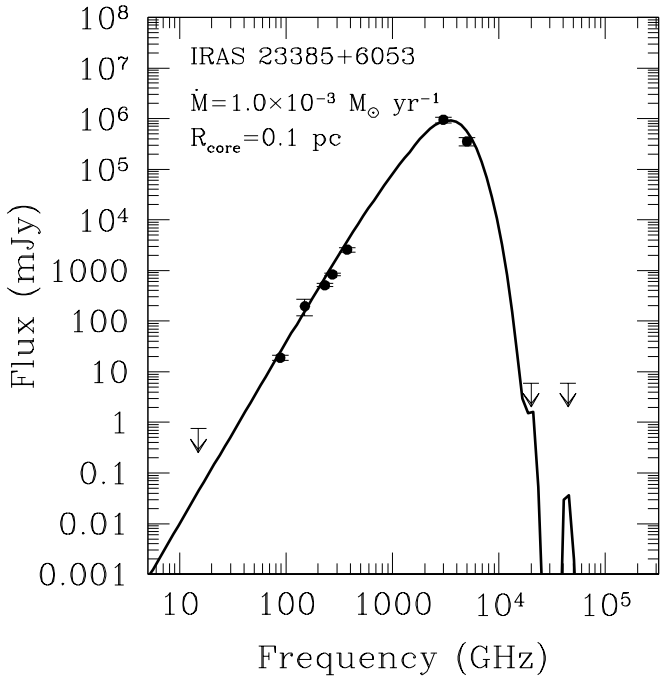
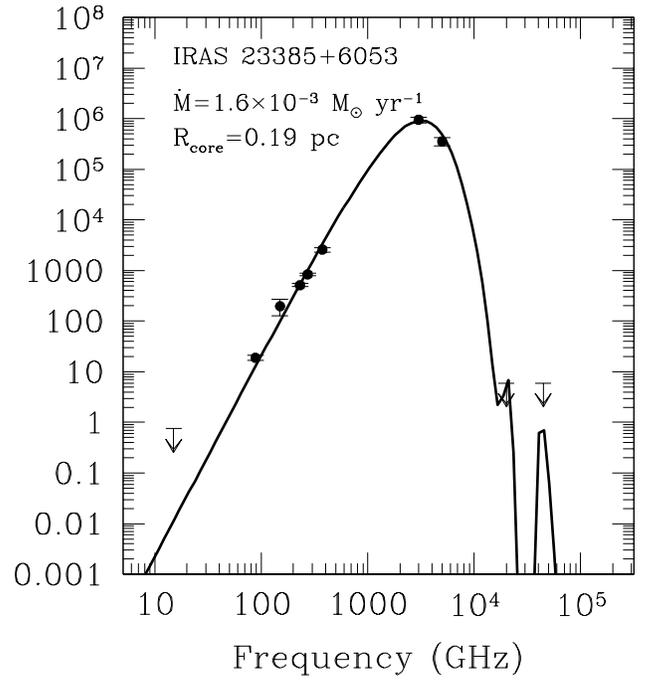
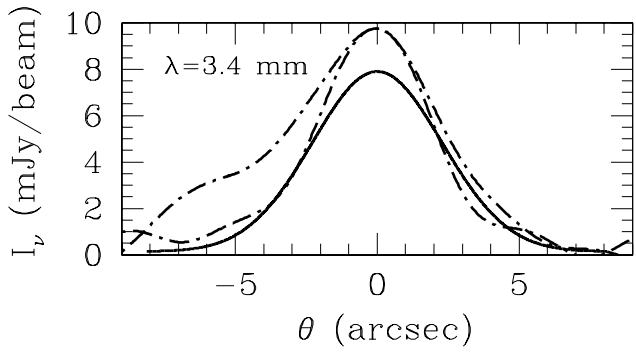
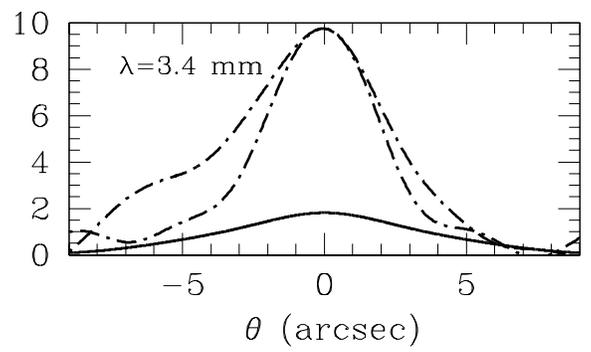
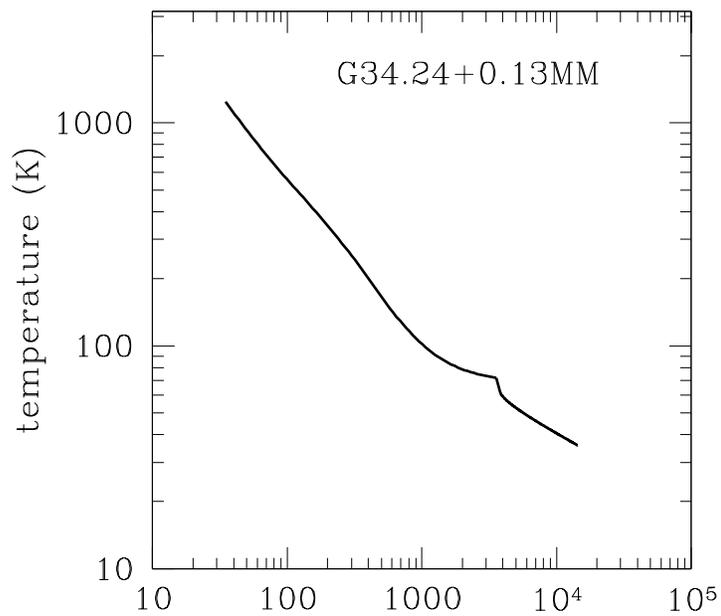
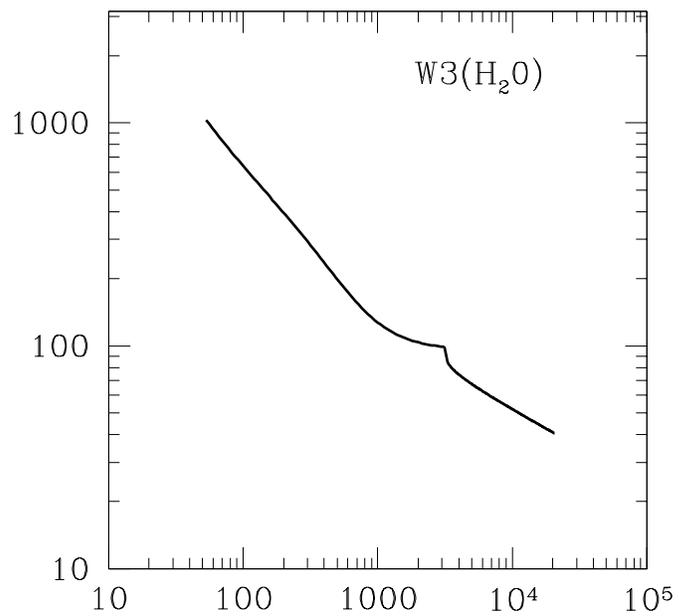
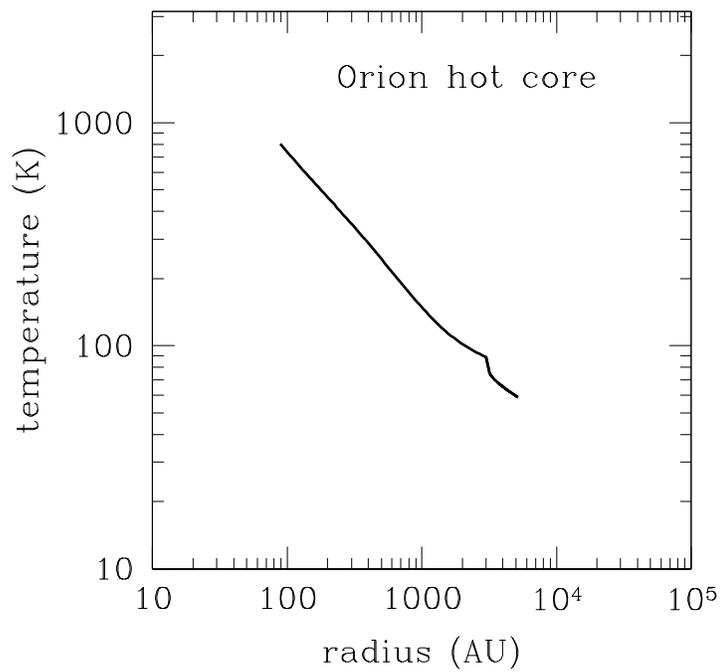
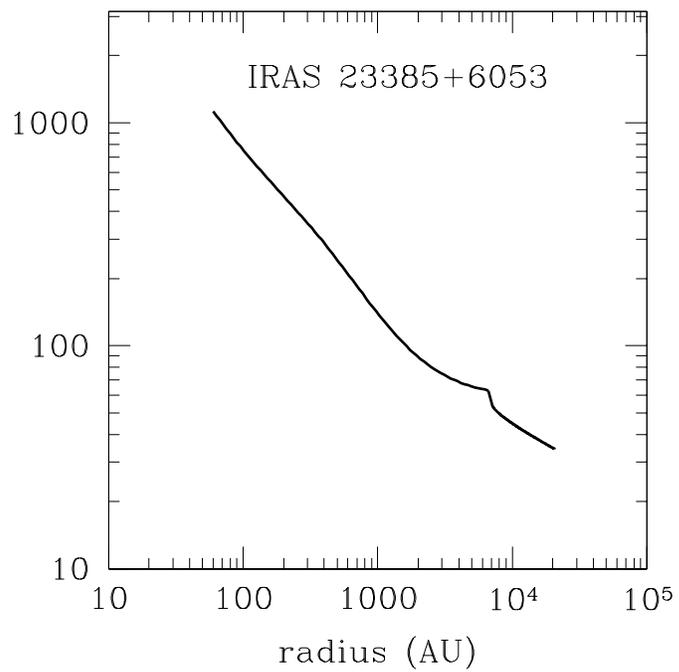
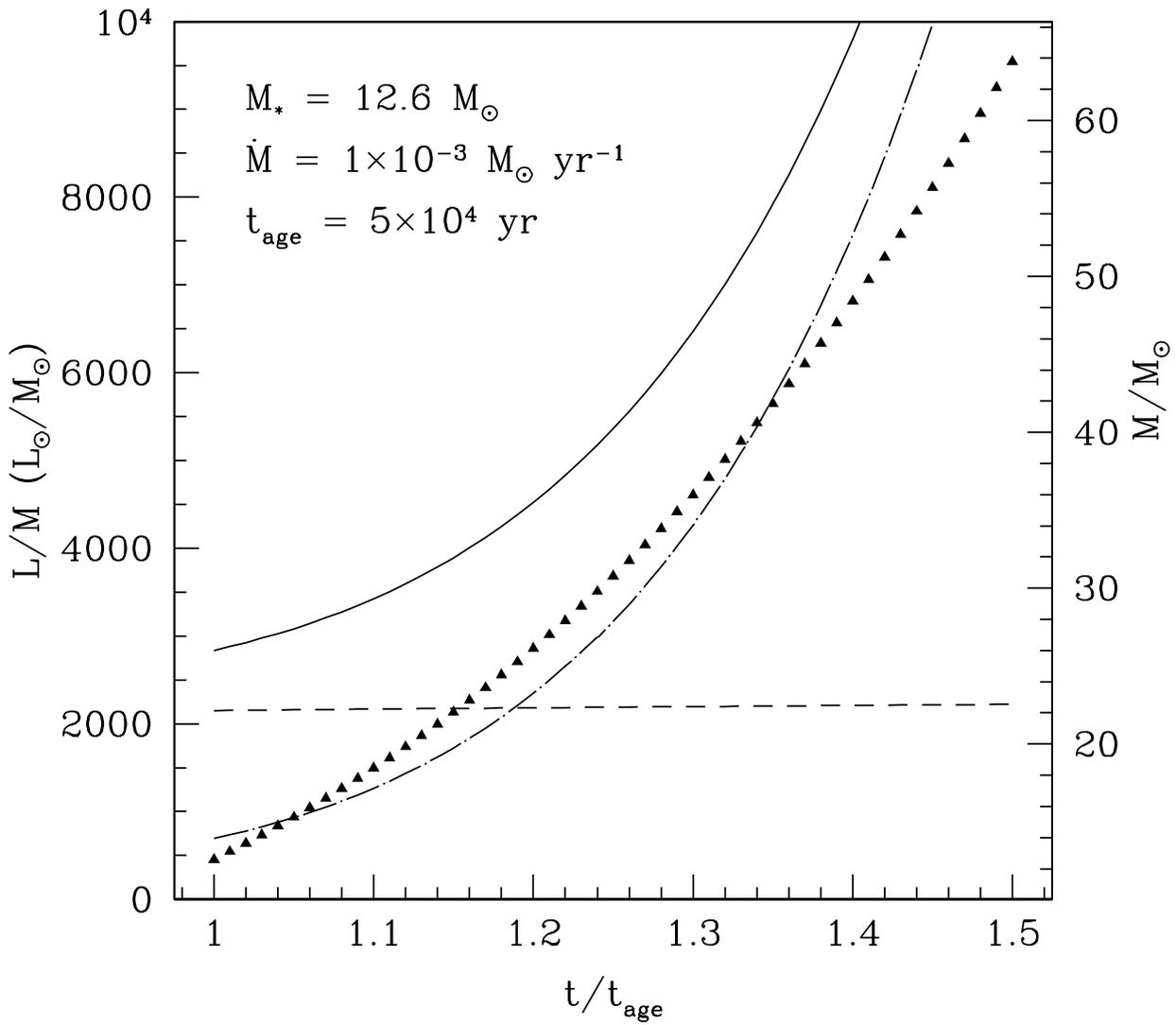
\end{document}